   \documentstyle[twocolumn,aps,prb,epsf]{revtex}

\def\e{\epsilon}         \def\s{\sigma}       	   	
\def\l{\lambda}	         \def\t{\tau}					
\def\D{\Delta}	         \def\W{\Omega}		   	
\def\k{{\bf k}}          \def\j{{\bf j}}				
\def\S{{\cal S}}	 \def\Hso{{\cal H}_{\rm so}}  
                  
\def\RT{R_{\rm T}}	 \def\GT{G_{\rm T}}	   \def\VT{V}
\def\DR{  R} 	         \def\AJ{A_{\rm J}}        \def\AN{A_{\rm N}}	   
\def\tGT{\tilde{G}_{\rm T}}	
\def\RN{{\cal R}_{\rm N}}\def\RF{{\cal R}_{\rm F}}			
    
\def\lN{\l_{\rm N}}	 \def\lF{\l_{\rm F}}	   \def\lS{\l_{\rm S}}
\def\sN{\s_{\rm N}}      \def\sF{\s_{\rm F}}        	 
     \def\muF{\mu_{\rm F}}
\def\muN{\mu_{\rm N}}    \def\dmuN{\delta\mu_{\rm N}} 
\def\rN{\rho_{\rm N}}    \def\rF{\rho_{\rm F}}     \def\sN{\s_{\rm N}}	
\def\dN{d_{\rm N}}       \def\dF{d_{\rm F}}     
\def\wN{w_{\rm N}}       \def\wF{w_{\rm F}}     
\def\/{\over}            \def\us{\uparrow}         \def\ds{\downarrow}	
	 	   
\def\[{\left[}           \def\]{\right]}	   
\def\({\left(}           \def\){\right)}	   
\def\<{\langle}          \def\>{\rangle}	   
\def\kT{k_{\rm B}T}	 \def\tS{\tau_{s}}         \def\tsf{\tau_{\rm sf}}
    
\def\Xs{\chi_s}		 \def\pF{p_{\rm F}}	   
\def\PJ{P_{\rm J}}       \def\dt{\partial t}	   
\def\d{\partial}         \def\h{\hbar}

\begin{document}
  \twocolumn[\hsize\textwidth\columnwidth\hsize\csname
  @twocolumnfalse\endcsname
\draft

\title{Spin Injection and Detection in Magnetic Nanostructures}

\author{S. Takahashi and S. Maekawa}

\address{Institute for Materials Research, Tohoku University, Sendai 980-8577, Japan}

\date{August 21, 2002}
\maketitle
\widetext

\begin{abstract}
We study theoretically the spin transport in a nonmagnetic metal
connected to ferromagnetic injector and detector electrodes.
We derive a general expression for the spin accumulation signal which
covers from the metallic to the tunneling regime.  This enables us to
discuss recent controversy on spin injection and detection experiments.
Extending the result to a superconducting device, we find that
the spin accumulation signal is strongly enhanced by opening
of the superconducting gap since a gapped superconductor is a
low carrier system for spin transport but not for charge.
The enhancement is also expected in semiconductor devices.
\end{abstract}


  \vskip1pc]
\narrowtext



There has been considerable interest recently in spin transport
in magnetic nanostructures. \cite{book}  The spin polarized electrons
injected from ferromagnets (F) into nonmagnetic materials (N) such as a
normal metal, semiconductor, and superconductor create
nonequilibrium spin accumulation in N.
\cite{aronov,johnson,jedema1,jedema2,otaniJMMM,ohno,fiederling,vasko,dong,chenPRL88,brataas}
The efficient spin injection, accumulation, and transport are central
issues to be explored in manipulating the spin degree of freedom of
the electron.
Johnson and Silsbee \cite{johnson}
have demonstrated that the injected spins penetrate
into N over the spin-diffusion length of $\mu$m scale using the spin
injection and detection techniques in F1/N/F2 trilayer structures.
Very recently, Jedema {\it et al.} have made a permalloy/copper/permalloy
(Py/Cu/Py) structure \cite{jedema1} and observed spin
accumulation at room temperature.  Subsequently, they have shown that
the efficiency of spin injection and accumulation is greatly improved
in a cobalt/aluminum/cobalt (Co/I/Al/I/Co) structure with tunnel
barriers (I). \cite{jedema2}

In this paper, we study the spin injection and detection in a device of
F1/N/F2 structure by taking into account the geometry of nonlocal
measurement. \cite{jedema1,jedema2}  By proper modeling of the system in
the diffusive transport regime, we derive an analytical expression for the
spin accumulation signal which covers from the metallic to the tunnel
regime.  
A controversial issue on the analysis of spin accumulation has been
raised in the structures of metallic contacts. \cite{johnsonNature}
We discuss the issue based on the present analytical expression.
Extending the result to the device containing a superconductor,
we find that the spin signal is greatly enhanced by opening
of superconducting gap.
Large spin signals are also expected in semiconductor devices.


We consider a spin injection and detection device consisting of 
a nonmagnetic metal N connected to ferromagnets of injector F1
and detector F2 as shown in Fig.~\ref{fig1}.
F1 and F2 are the same ferromagnetic films of width $\wF$
and thickness $\dF$, and are separated by distance $L$.
N is a normal metal film of width $\wN$ and thickness $\dN$.
The magnetizations of F1 and F2 are aligned either parallel or
antiparallel.
Since the spin-diffusion length $\lN$ of N
($\l_{\rm Cu} \sim 1\mu$m, \cite{jedema1}
$\l_{\rm Al} \agt 1\mu$m \cite{johnson,jedema2,otaniJMMM})
is much larger than the length $\lF$ of F
($\l_{\rm Py} \sim 5$~nm \cite{steenwyk}), we consider the device
having dimensions of $\lF \ll (\dN, \dF) \ll (\wN, \wF) \ll \lN$.
This situation, which corresponds to the experimental geometry,
  \cite{jedema1,jedema2}
facilitates a description for the spin and charge transport
in the device.

The electrical current $\j_\s$ for spin channel $\s$ is driven by
the electric field ${\bf E}$ and the gradient of the carrier density
deviation $\delta n_\s$ from equilibrium:
   $\j_\s = {\s_\s } {\bf E} - e D_\s \nabla \delta n_\s$,	
where $\s_\s$ and $D_\s$ are the electrical conductivity and the
diffusion constant.
Making use of $\delta n_\s=N_\s\delta\e_\s$ and $\s_\s=e^2 N_\s D_\s$
($N_\s$ is the density of states in the spin subband and
$\delta\e_\s$ is the shift in the chemical potential of carriers
from its equilibrium value) gives
   $\j_\s = -{(\s_\s/e) } \nabla \mu_\s$,	
where $\mu_\s=\e_\s+e\phi $ is the electrochemical potential (ECP)
and $\phi$ the electric potential.
The continuity equations for charge and spin in the steady state are
$ \nabla \cdot \(\j_\us + \j_\ds \) = 0$ and
$ \nabla \cdot \(\j_\us - \j_\ds \)  = - e{\delta n_\us /\t_{\us\ds}} 		
  	+ e{\delta n_\ds /\t_{\ds\us}}$,
where $\t_{\s\s'}$ is the scattering time of an electron from
spin state $\s$ to $\s'$.
Using these equations and detailed balancing
$N_\us/\t_{\us\ds}=N_\ds/\t_{\ds\us}$,
one obtains \cite{johnson,vanson,valet,fertlee,hershfield,rashba}
\begin{eqnarray}				
  \nabla^2 \(\s_\us\mu_\us + \s_\ds\mu_\ds\) = 0,
  \ \ \ \ \ \label{eq:ddr1}	\\		
  \nabla^2 \(\mu_\us - \mu_\ds\)		
    = \l^{-2} \(\mu_\us - \mu_\ds\),	
   \label{eq:ddr2}				
\end{eqnarray}					
with the spin-diffusion length $\l = \sqrt{D \tsf}$, where
$\tsf^{-1}=
{1\/2}(\t_{\us\ds}^{-1}+\t_{\ds\us}^{-1})$
and
$D^{-1} = (N_\us D^{-1}_\ds+N_\ds D^{-1}_\us)/(N_\us+N_\ds)$. 
The material parameters in N are spin-{\it independent}:
\vfill
\begin{figure}[thb]					
  \epsfxsize=0.84\columnwidth				
  \centerline{\hbox{\epsffile{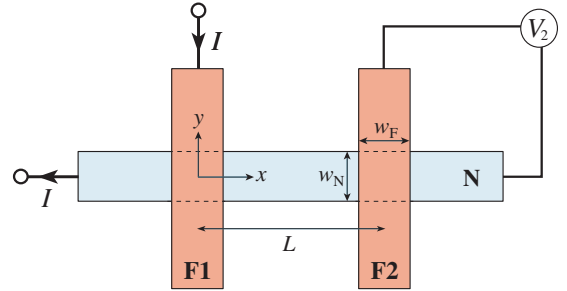}}}		
  \vskip 0.40cm
  \caption{						
Basic structure of a spin injection and		
detection device.  The bias current $I$ flows from F1	
to the left side of N.  The spin accumulation at distance
$L$ is detected by measuring the spin-dependent voltage	
$V_2$ between F2 and N.					
  }   \label{fig1}					
\end{figure}						
\noindent
$\sN^\us=\sN^\ds={1\/2}\sN$, $D_\us=D_\ds$, 
etc., and those in F spin-{\it dependent}:
$\sF^\us \ne \sF^\ds$ ($\sF=\sF^\us+\sF^\ds$),
$D_\us \ne D_\ds$, etc.

We employ a simple model for the interfacial currents of the junctions. 
The distribution of the interfacial spin currents is uniform
over the contact area $\AJ=\wF\wN$ since the $\lN$ is much
longer than $\wF$ and $\wN$, and ECP has a discontinuous drop
at the interface of junction $i$ ($i=1,2$) associated with the
interface resistance $R_i$.
\cite{johnson,valet,fertlee,hershfield,rashba}
We neglect the interfacial spin-flip scattering
\cite{fertlee,rashba} for simplicity.
The interfacial current $I_i^{\s}$
across the interface ($z=0$) is given by
    $I_i^{\s} = (G_i^\s/e) \( \muF^\s|_{z=0^+} - \muN^\s|_{z=0^-} \)$,
where $G_i^\s$ is the interface conductance 
($G_i=G_{i}^{\us}+G_{i}^{\ds} = R_i^{-1}$).  
In the transparent contact ($G_i \rightarrow \infty$) the ECPs are
continuous at the interfaces, while in the tunneling junction
the discontinuity in ECP is much larger
than the spin splitting in ECP.
The interfacial charge and spin currents are
$I_i=I_{i}^{\us}+I_{i}^{\ds}$ and $I_i^{s}=I_{i}^{\us}-I_{i}^{\ds}$.


When the bias current $I$ flows from F1 to the left side of N ($I_1=I$)
and no charge current through the F2/N junction ($I_2=0$),
the solutions for ECPs that satisfy Eqs.~(\ref{eq:ddr1}) and
(\ref{eq:ddr2}) are constructed as follows.
In the N electrode whose thickness and contact dimensions are much
smaller than $\lN$, $\muN^\s$ varies only in the $x$ direction:
   $\muN^\s(x) = {\bar \muN} + \s \dmuN$,	
where 
${\bar \muN} = -{(eI/\sN)}x$ for $x<0$, ${\bar \muN} = 0$
for $x>0$, and 
$\dmuN = a_1 e^{-|{x}|/\lN} + a_2 e^{-|x-{L}|/\lN}$
with the $a_1$-term being the ECP shift due to spin injection
from F1 at $x=0$, and the $a_2$-term being the feedback shift
due to the presence of F2 at $x=L$.
The spin current
$j_{s}=j_\us-j_\ds$ flows in the $x$ direction according to
$j_{ s}=-(\sN/e) \nabla \dmuN$.   
The continuity of the spin current at junction $i$ yields
$I_i^{s} = 2(\sN\AN/e\lN) a_i$, where  
$\AN=\wN\dN$ is the
cross-sectional area of N.
Note that only the spin current flows in the region of $x>0$
and no charge current there.

In the F1 and F2 electrodes whose thickness and contact dimensions are
much larger than $\lF$, the spin splitting of $\muF^\s$ decays quickly
along the $z$ direction, so the solution has the form near the
interface ($0 < z \alt \lF$):
   $\muF^\s(z) = {\bar \muF} + \s b_i\(\sF/{\sF^\s}\) e^{-z/\lF}$,
where ${\bar \muF} = -{(eI/\sF\AJ)}z + eV_1$ in F1 and
${\bar \muF} = eV_2$ in F2, $V_1$ and $V_2$ being the voltage drops
  $({\bar \muF}-{\bar \muN})/e$
at the interfaces of junctions 1 and 2, respectively.  
The continuity of the spin currents at the junctions leads to
 $I_1^s=\pF I-2(\sF\AJ/e\lF) b_1$ and 
 $I_2^{s}=-2(\sF\AJ/e\lF) b_2$, where 
$\pF={(\sF^\us-\sF^\ds)/(\sF^\us+\sF^\ds)}$
is the current polarization of F1 and F2.
The constants $a_i$, $b_i$, and $V_i$
are determined by the continuity condition for the spin and
charge currents at the interfaces.


The spin-dependent voltage $V_2$ detected at F2, i.e., the potential
difference between the right side of N electrode and the F2 electrode,
is given by
  \begin{eqnarray}					
    {V_2/I} &=& \pm {2\RN} e^{-{L\/\lN}}
    \prod_{i=1}^2 \({\PJ {R_i\/\RN} \/1-\PJ^2} +
    {\pF {\RF\/\RN} \/1-\pF^2}  \) 
     \cr  &\times&
    \[ \prod_{i=1}^2 \( 1 + {2 {R_i\/\RN} \/1-\PJ^2}
    +{2 {\RF\/\RN} \/ 1-\pF^2} \) - e^{-{2L\/\lN}}\]^{-1},
     \label{eq:V2}			
  \end{eqnarray}					
where signs ``{+}" and ``{$-$}" correspond to the parallel (P)
and antiparallel (AP) alignments of magnetizations, 
$\RN = \rN\lN/\AN$ and $\RF = \rF\lF/\AJ$
represent the resistances of N and F with cross-sections $\AN$ and $\AJ$
and lengths $\lN$ and $\lF$, respectively, and
$\PJ = {|G_i^\us-G_i^\ds|/G_i}$ is the interfacial current polarization,
and $\rN = \sN^{-1}$ and $\rF=\sF^{-1}$ are the resistivities.
The spin accumulation signal is detected as 
the voltage change $V_s=(V_2^{\rm P}-V_2^{\rm AP})=2|V_2|$ or
the resistance change $\DR_s=V_s/I$ when the magnetizations
change from the P to AP alignment.

The spin accumulation signal $\DR_s$ depends on whether each junction
is a metallic contact or a tunnel junction.
Since $\RF/\RN$ $\sim 0.01$ for the typical values ($\rF/\rN$ $\sim 10$,
$\lF/\lN \sim 0.01$, and $\AN/\AJ \sim 0.1$), \cite{jedema1}
we have the following limiting cases; 
When both junctions are transparent contact ($R_1,R_2 \ll \RF$), 
we have \cite{jedema1,fertlee,hershfield,rashba}
  \begin{eqnarray}			
    \DR_s =  {4\pF^2 \/ (1-\pF^2)^2} 	
    \RN{\(\RF\/\RN\)^2}			
     {e^{-{L/\lN}}\/ 1-e^{-{2L/\lN}} }.	
     \label{eq:V2-mm}			
  \end{eqnarray}			
When one of the junctions is a transparent contact and the other
is a tunnel junction, i.e.,
 ($R_1 \ll \RF \ll \RN \ll R_2$) or
 ($R_2 \ll \RF \ll \RN \ll R_1$), we have
  \begin{eqnarray}			
    \DR_s =  {2\pF\PJ \/ (1-\pF^2)} \RN	
     {\(\RF\/\RN\)} {e^{-{L/\lN}} } .	
     \label{eq:V2-tm}			
  \end{eqnarray}			
When both junctions are tunneling junctions ($R_1,R_2 \gg \RN$),
we have \cite{johnson,jedema2}
  \begin{eqnarray}		
    \DR_s = \PJ^2 \RN e^{-{L/\lN}}.	
     \label{eq:V2-tt}		
  \end{eqnarray}		
Note that $\DR_s$ in the above limiting cases is independent of $R_i$.
Equations~(\ref{eq:V2-mm})-(\ref{eq:V2-tt}) indicate that 
the resistance mismatch factor $(\RF/\RN)$ is removed
systematically when a transparent contact is replaced with a tunnel
junction. \cite{rashba,fertPRB64,schmidt,mismatch}
Thus, the maximum spin signal is achieved when all the
junctions are tunnel junctions.

\begin{figure}					
  \epsfxsize=0.88\columnwidth			
  \centerline{\hbox{\epsffile{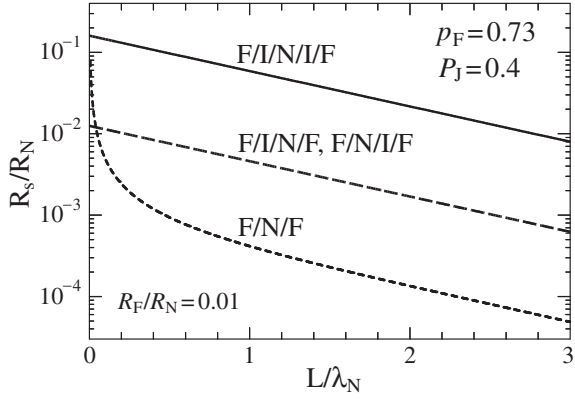}}}	
  \vskip 0.40cm
  \caption{					
Spin accumulation signal $\DR_s$ vs distance	
$L$ between F1 and F2.	 Solid line:~F/I/N/I/F.	
Long-dashed line:~F/I/N/F and F/N/I/F. 		
Short-dashed line:~F/N/F.  
  }						
  \label{fig2}						
\end{figure}

Figure \ref{fig2} show the spin accumulation signal
$\DR_s$ in Eqs.~(\ref{eq:V2-mm})-(\ref{eq:V2-tt})
for $\pF=0.73$, \cite{steenwyk} $\PJ=0.4$,
\cite{soulen} and $\RF/\RN = 10^{-2}$. \cite{jedema1} 
We see that $\DR_s$ increases by one order of magnitude 
by replacing a transparent contact with a tunnel barrier.
The value $\RN=3~\W$ \cite{Cu}
taken from the Py/Cu/Py structure yields
$\DR_s = 1{\,}{\rm m}\W$ at $L=\lN$.   
If one takes into account the cross-shaped Cu, \cite{jedema1}
one expects one-third of the above value, which is in reasonable
agreement with the experimental value $0.1{\,}{\rm m}\W$. \cite{jedema1} 
In the Co/I/Al/I/Co structure, $\RN=3{\,}\W$ is estimated
\cite{Al} and 
$\DR_s = 100{\,}{\rm m}\W$ is obtained at $L=\lN$,
which is 10 times larger than the experimental value
$10{\,}{\rm m}\W$. \cite{jedema2}
This discrepancy may be attributed to the reduction in $\PJ$
due to the spin-flip scattering at the barriers.
\cite{fertlee,rashba}

A question arises on whether the contacts of metallic F1/N/F2
structures is transparent ($R_i\ll \RF$) \cite{jedema1} or 
tunneling-like ($R_i \gg \RN$). \cite{johnsonNature,johnson_bayers} 
The experimental values 
\noindent
of Py/Cu 
($R_i\AJ \sim 5 \times 10^{-12} {\,}\W$cm$^2$,\cite{steenwyk}
$\rF \sim 10^{-5}{\,}\W$cm, \cite{jedema1} and $\lF \sim 5$nm \cite{steenwyk})
yields $R_i \sim \RF$, which is strictly speaking neither transparent
or tunneling-like.  However, the values of $\DR_s$ for $R_i = \RF$
calculated from Eq.~(\ref{eq:V2}) are about 2 times larger than those
for the transparent case in
Fig.~\ref{fig2}, indicating that the Py/Cu/Py structure lies on
the verge of transparent regime.   
However, depending on sample fabrication processes, there will be
cases that belong to the intermediate
regime ($\RF \ll R_i \ll \RN$), for which one should use
  \begin{eqnarray}			
    \DR_s =  {4\PJ^2 \/ (1-\PJ^2)^2}	
    \RN{\(R_1R_2\/\RN^2\)}		
     {e^{-{L/\lN}}\/ 1-e^{-{2L/\lN}} }.	
     \label{eq:V2-mid}			
  \end{eqnarray}			
If $R_i \sim \RN$, then $\DR_s$ is close to the values of tunneling case,
so that the contacts of $R_i \agt \RN$ belong to the tunneling regime.


The spin injection into a superconductor (S) is of great
interest from basic and practical points of views.
We show that S becomes a low-carrier system for spin transport
by opening of the superconducting gap $\D$ and the resistivity
of the spin current increases below the superconducting critical
temperature $T_c$.
In the tunneling device of F1/I/S/I/F2, 
the spin signal would increase due to the
increase of $\RN$ below $T_c$ [see Eq.~(\ref{eq:V2-tt})].  
 Therefore, we investigate in detail how the spin signal
 is enhanced by opening of $\D$.
In the following, we consider the situation where the spin
splitting of ECP, the maximum of which is 
$\dmuN(0) \sim {1\/2}e\PJ \RN I $, is smaller than $\D$, i.e.,
$I < {2\D/(e\PJ\RN)}$, for which the suppression of $\D$ due to spin
accumulation can be neglected. \cite{takahashiPRL82}
We also neglect charge imbalance created by injection
of QP charge into S, which originates from the conversion of
injected QPs into condensate, and produces the excess voltage
due to charge accumulation at F2. \cite{clarkePRL28}
However, the effect is spin-independent and does not contribute to $\DR_s$.

In the superconducting state, the equation for the spin splitting
$(\mu_\us-\mu_\ds)$ is the same as Eq.~(\ref{eq:ddr2}) with
$\lN$ in the normal-state, \cite{yamashitaPRB}
which is intuitively understood as follows. 
Since the dispersion curve of the QP excitation energy is given by
$E_k=\sqrt{\e^2_k+\D^2}$ with one-electron energy $\e_k$, the QP's
velocity $\tilde{v}_k=(1/\h)(\d E_k/\d k) = (|\e_k|/E_k) v_{k}$
is slower by the factor $|\e_k|/E_k$ compared with the normal-state
velocity $v_k (\approx v_{\rm F})$.
By contrast, the impurity scattering time 
$\tilde{\tau}_{\s\s'}=(E_k/|\e_k|)\tau_{\s\s'}$
\cite{bardeen} is longer by the inverse of the factor.
Then, the spin-diffusion length
$\lS=(\tilde{D}\tilde{\t}_{sf})^{1/2}$ in S with
 $\tilde{D}={1\/3} \tilde{v}^2_{k}\tilde{\t}_{\rm imp}$
and 
 $\tilde{\t}^{-1}_{\rm imp}=\sum_{\s'} \tilde{\tau}^{-1}_{\s\s'} $
results in $\lS=\sqrt{{D}{\t}_{sf}}=\lN$ owing to the cancellation
of the factor $|\e_k|/E_k$.
Consequently, the spin splitting in S has the same form of solution
as in N.

Utilizing the so-called semiconducting picture for electron tunneling
between F and S, the charge and spin currents across junction 1
are calculated as
  $I = \tGT V$ and
  $I^{s}_1 = \PJ \tGT \VT$
 at low bias $\VT$ ($ = V_1 < \D$), \cite{takahashiJMMM}
and those across junction 2 are 
  $I_2=\tGT [V_2-P_2 \dmuN(L)/e]=0$ and
  $I^{s}_2 = P_2 \tGT V_2$.
Here, $P_2$ takes $\PJ$ ($-\PJ$) for the P (AP) alignment, 
$\tGT=\Xs(T)\GT$ is the tunnel conductance in the
superconducting state, and $\Xs(T)$ is the Yosida function \cite{Xs}
which represents the reduction of the tunnel conductance by opening
of $\D$ below $T_c$.

The spin accumulation in S is determined by balancing
the spin injection rate with the spin-relaxation rate:
$I_i^{s} + e(\d\S_i/\dt)_{\rm sf}=0$,
where $\S_i$ is the total spins accumulated in S by spin injection
through junction $i$.  At low temperatures the spin relaxation is
dominated by the spin-flip scattering via the spin-orbit interaction
$\Hso$ at nonmagnetic impurities or grain boundaries.
The scattering matrix elements of $\Hso$ over quasiparticle states
$|\k\s\>$ with momentum $\k$ and spin $\s$ has the form:
$ \<\k'\s'|\Hso|\k\s\> \propto i \(u_{k'}u_{k} -v_{k'}v_{k}\)
[ {\vec\s}_{\s'\s}\cdot ({\k}\times{\k'}) ]$,
where 
$\vec{\s}$ is the Pauli spin matrix and
  $u_k^2 = 1-v_k^2 = {1\/2}\( 1+{\e_k/E_k}\)$.
Using the golden rule formula, \cite{takahashiJMMM}
we can calculate $ (\d\S_{i}/\dt)_{\rm sf}$
and obtain $I^s_i = [2 f_0(\D)/(e\RN)] a_i$, where
$2f_0(\D)$ represents the QP populations and
$f_0(\D)=1/[\exp({\D/\kT})+1]$.

From the matching condition of the spin currents across the barriers,
we obtain the spin signal $\DR_s$ in the superconducting state
  \begin{eqnarray}			
    \DR_s = {V_s/I}
          ={1\/2f_0(\D)}\PJ^2	
     \RN {e^{-{L/\lN}} }.		
     \label{eq:V2-I}			
  \end{eqnarray}			
If the $I-V$ characteristics, $I=\Xs(T)\VT/\RT$, is used,
  \begin{eqnarray}			
    {V_s/\VT} = {\Xs(T)\/2f_0(\D)}\PJ^2 
    {\RN\/\RT} {e^{-{L/\lN}} } .	
     \label{eq:V2-V1}			
  \end{eqnarray}			
The above results are obtained from those of the normal state by the
scaling $\rN \rightarrow \rN/[2f_0(\D)]$ and $\RT \rightarrow \RT/\Xs(T)$.
Equation~(\ref{eq:V2-I}) is interpreted as follows: The spin-current 
density in SC is given by 
$j_{s} = -({\sN/e})$ $2f_0(\D) \nabla \dmuN$,
\cite{takahashiPRL88} where the effective conductivity 
$2f_0(\D)\sN$ decreases due to the decrease of QP populations
by opening the gap $\D$ below $T_c$.  The boundary condition that
the injected spin current $\PJ I$ is equal to $2j_s(0^+)\AN$ yields 
$\dmuN \approx \{e\PJ I\RN/[2 f_0(\D)]\}e^{-|x|/\lN}$.
The decrease of the effective conductivity is compensated
by the increase of $\dmuN$ to maintain the same spin injection
in the constant $I$, and therefore $\DR_s$ increases as 
${1\/2}f^{-1}_0(\D)$ below $T_c$.
Note that the $T$-dependent factor in Eq.~(\ref{eq:V2-V1}) is 
\begin{figure}						
  \epsfxsize=0.83\columnwidth				
  \centerline{\hbox{\epsffile{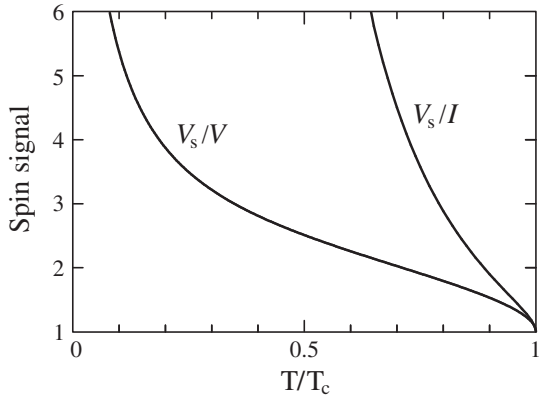}}}		
  \vskip 0.40cm
  \caption{						
Temperature dependence of the spin signals $\DR_s=	
V_s/I$ and $V_s/\VT$ in a F/I/S/I/F structure.  	
The values of the spin signal are normalized to those at
the superconducting critical temperature $T_c$.		
   }							
  \label{fig3}						
\end{figure}						
\noindent
the same as that in the spin-relaxation
time $\tS = \[{\Xs(T)/2f_0(\D)}\]\tsf$, \cite{yafet} which is
derived from $\(\d\S / \dt\)_{\rm sf} = - {\S / \tS}$.


Figure~\ref{fig3} shows the temperature dependence of
$\DR_s=V_s/I$ and $V_s/V$.  The values are normalized
to those at $T_c$.   The strong increase of $V_s/I$ reflects
the $T$-dependence of the resistivity of the spin current below $T_c$.
The signal $V_s/V$ increases with the same $T$-dependence as $\tS(T)$,
indicating that the spin-relaxation time in S is directly
obtained by measuring $V_s$ vs $T$ at constant $V$.
To test these predictions, it is highly desirable to measure $V_s$
of Co/I/Al/I/Co structures \cite{jedema2} by lowering $T$ below $T_c$.

A large enhancement of spin signals is also expected in degenerate
semiconductors, because the resistivity is much larger compared with
normal metals and the spin-diffusion length is relatively long.  
In degenerate semiconductors, the spin current is given by
   $j_{s} = -\mu_m n_c \nabla_x (\mu_\us-\mu_\ds)$,	
where $\mu_m$ is the mobility and $n_c$ the carrier concentration. 
For Si-doped GaAs with $n_c=10^{18}$cm$^{-3}$ and 
$\mu_m=2 \times 10^{3}$cm$^2$/Vs at room temperature, \cite{kikkawa}
$\rN = 1/(e\mu_m n_c) = 0.1{\,}\W$cm.
For (Mn,Ga)As, $\rF = 0.01 \sim 0.1 {\,}\W$cm. \cite{matsukura}
It follows from Eqs.~(\ref{eq:V2-tm}) and (\ref{eq:V2-tt}) that
$\DR_s \propto \rF$ for a (Ga,Mn)As/I/GaAs($n$-type)/(Ga,Mn)As device
and $\DR_s \propto \rN$ for a F1/I/GaAs($n$-type)/I/F2 device. 
Therefore, we expect that $\DR_s$ is larger by several orders of
magnitude than that of metal case.
This result is promising for applications for spintronic devices.

In summary, we have studied the spin injection and detection in the
F1/N/F2 structure, and derived an expression for the spin accumulation
signal which covers from the metallic to the tunneling regime.  
This enables us to resolve the recent controversy of spin injection
and detection experiments.  
 Extending the result to a superconducting device, we have found
that the signal is strongly enhanced below $T_c$, because superconductors 
become a low carrier system for spin transport by opening of the gap
and a larger spin splitting is required for carrying the same spin current.  
 Our finding can be tested in superconducting devices such as
Co/I/Al/I/Co by lowering temperature below $T_c$.
A large spin accumulation signal is also expected in semiconductor devices.

This work is supported by a Grant-in-Aid for Scientific Research
from MEXT and CREST. 




\begin{thebibliography}{9}
\small

\bibitem{book}
{\it Spin Dependent Transport in Magnetic Nanostructures}, edited by
S. Maekawa and T. Shinjo (Taylor and Francis, London and New York, 2002).

\bibitem{aronov}
A. G. Aronov, JETP Lett. {\bf 24}, 32 (1976); JETP {\bf 44}, 193 (1976); 
Sov. Phys. Semicond. {\bf 10}, 698 (1976).

\bibitem{johnson}
M. Johnson and R.~H. Silsbee, Phys. Rev. Lett. {\bf 55}, 1790 (1985);
{\it ibid.} {\bf 60}, 377 (1988);
{\it ibid.} {\bf 70}, 2142 (1993).

\bibitem{jedema1}
F.J. Jedema {\it et al.}, Nature (London) {\bf 410}, 345 (2001); cond-mat/0207641.

\bibitem{jedema2} 
F.J. Jedema {\it et al.}, Nature (London) {\bf 416}, 713 (2002).

\bibitem{otaniJMMM}
Y. Otani {\it et al.}, J. Magn. Magn. Mater.
{\bf 239}, 135 (2002).

\bibitem{ohno}
Y. Ohno  {\it et al.}, Nature (London) {\bf 402}, 790 (1999).

\bibitem{fiederling}
R. Fiederling {\it et al.}, Nature (London) {\bf 402}, 787 (1999). 

\bibitem{vasko}
V.~A. Vas'ko {\it et al.}
Phys. Rev. Lett. {\bf 78}, 1134 (1997).

\bibitem{dong}
Z.~W. Dong {\it et al.}  
Appl. Phys. Lett. {\bf 71}, 1718 (1997).

\bibitem{chenPRL88}
C. D. Chen et al., Phys. Rev. Lett. {\bf 88}, 047004 (2002).

\bibitem{brataas}
A. Brataas {\it et al.}, Phys. Rev. Lett. {\bf 84}, 2481 (2000).

\bibitem{johnsonNature}
M. Johnson {\it et al.}, Nature {\bf 416}, 809 (2002).

\bibitem{steenwyk}
S.D. Steenwyk {\it et al.}, J. Magn. Magn. Mater. {\bf 170}, L1 (1997);
S. Dubois {\it et al.}, Phys. Rev. B {\bf 60}, 477 (1999).

\bibitem{vanson}
P. C. van Son {\it et al.} Phys. Rev. Lett. {\bf 58}, 2271 (1987).

\bibitem{valet}
T. Valet and A. Fert, Phys. Rev. B {\bf 48}, 7099 (1993).

\bibitem{fertlee}
A. Fert and S.F. Lee,  Phys. Rev. B {\bf 53}, 6554 (1996).

\bibitem{hershfield}
S. Hershfield and H.L. Zhao, Phys. Rev. B {\bf 56}, 3296 (1997).

\bibitem{rashba}
E.I. Rashba, Phys. Rev. B {\bf 62}, R16267 (2000);
cond-mat/0206129.

\bibitem{fertPRB64}
A. Fert and H. Jaffr\`es, Phys. Rev. B {\bf 64}, 184420 (2001). 

\bibitem{schmidt}
G. Schmidt  {\it et al.}, Phys. Rev. B {\bf 62}, R4790 (2000).

\bibitem{mismatch}
The mismatch arises from those in length and area
($\lN \gg \lF$, $\AJ \gg \AN$), not in the resistivity
as in Refs.~[19-21].

\bibitem{soulen}
R. J. Soulen Jr. {\it et al.}, Science {\bf 282}, 85 (1998).

\bibitem{johnson_bayers}
M. Johnson and J. Bayers (unpublished).

\bibitem{Cu}
$\rho_{\rm N}=1.4{\,}\mu\W$cm, $\l_{\rm N}=1{\,}\mu$m, 
and $A_{\rm N}=100\times 50{\,}$nm$^2$.

\bibitem{Al}
$\rho_{\rm N}=6{\,}\mu\W$cm, $\l_{\rm N}=0.65{\,}\mu$m, 
and $A_{\rm N}=250\times 50{\,}$nm$^2$.

\bibitem{takahashiPRL82}
S. Takahashi {\it et al.}, Phys. Rev. Lett. {\bf 82}, 3911 (1999).

\bibitem{clarkePRL28}
J. Clarke, Phys. Rev. Lett. {\bf 28}, 1363 (1972);
M. Tinkham and J. Clarke, Phys. Rev. Lett. {\bf 28}, 1366 (1972).

\bibitem{yamashitaPRB}
T. Yamashita {\it et al.}, Phys. Rev. B {\bf 65}, 172509 (2002).

\bibitem{bardeen}
J. Bardeen {\it et al.}, Phys. Rev. {\bf 113}, 982 (1959).

\bibitem{takahashiJMMM}
S. Takahashi {\it et al.}, J. Magn. Magn. Mater. {\bf 240}, 100 (2002).

\bibitem{takahashiPRL88}
S. Takahashi and S. Maekawa, 
Phys. Rev. Lett. {\bf 88}, 116601 (2002).

\bibitem{Xs} 
$\Xs(T) = 2 \int_\D^\infty {E\/\sqrt{E^2-\D^2}}[-{(\d f_0(E)/\d E)}] dE$,
whose aymptotic vaues are $ 1 - [7\zeta(3) /4\pi^2] \(\D/\kT\)^2$
near $T_c$ and $ (\pi\D/2\kT)^{1/2} \exp[{-\D/\kT}]$ well below $T_c$.


\bibitem{yafet}
Y. Yafet, Phys. Lett. A {\bf 98}, 287 (1983).

\bibitem{kikkawa}
J.M. Kikkawa and D.D. Awschalom, 
Phys. Rev. Lett. {\bf 80}, 4313 (1998).

\bibitem{matsukura}
F. Matsukura {\it et al.}  Phys. Rev. B {\bf 57}, 2037 (1998).


\end{thebibliography}
\end{document}